\documentclass[%
 reprint,
 superscriptaddress,
 amsmath,amssymb,
 aps,
 prb,
 floatfix,
 notitlepage]{revtex4-1}

\usepackage{graphicx}

\usepackage{hyperref}
\hypersetup{colorlinks = true, citecolor = blue, breaklinks = true}

\usepackage[version=4]{mhchem}

\usepackage{xr}
\begin{document}

\title{Conversion of \ce{La2Ti2O7} to \ce{LaTiO2N} via Ammonolysis: An \textit{ab-initio} Investigation}

\author{Chiara Ricca}
\affiliation{%
Department of Chemistry and Biochemistry, University of Bern, Freiestrasse 3, CH-3012 Bern, Switzerland 
}%
\affiliation{%
National Centre for Computational Design and Discovery of Novel Materials (MARVEL), Switzerland
}%

\author{Tristan Blandenier}
\affiliation{%
Department of Chemistry and Biochemistry, University of Bern, Freiestrasse 3, CH-3012 Bern, Switzerland 
}%

\author{Valérie Werner}
\affiliation{%
Department of Chemistry and Physics of Materials, University of Salzburg, Jakob-Haringer-Str. 2A, A-5020 Salzburg, Austria 
}%

\author{Xing Wang}
\affiliation{%
Department of Chemistry and Biochemistry, University of Bern, Freiestrasse 3, CH-3012 Bern, Switzerland 
}%

\author{Simone Pokrant}
\affiliation{%
Department of Chemistry and Physics of Materials, University of Salzburg, Jakob-Haringer-Str. 2A, A-5020 Salzburg, Austria 
}%

\author{Ulrich Aschauer}
\email{ulrich.aschauer@plus.ac.at}
\affiliation{%
Department of Chemistry and Biochemistry, University of Bern, Freiestrasse 3, CH-3012 Bern, Switzerland 
}%
\affiliation{%
National Centre for Computational Design and Discovery of Novel Materials (MARVEL), Switzerland
}%
\affiliation{%
Department of Chemistry and Physics of Materials, University of Salzburg, Jakob-Haringer-Str. 2A, A-5020 Salzburg, Austria 
}%

\date{\today}

\begin{abstract}
		Perovskite oxynitrides are, due to their reduced band gap compared to oxides,  promising materials for photocatalytic applications. They are most commonly synthesized from \{110\} layered Carpy-Galy (\ce{A2B2O7}) perovskites via thermal ammonolysis, i.e. the exposure to a flow of ammonia at elevated temperature. The conversion of the layered oxide to the non-layered oxynitride must involve a complex combination of nitrogen incorporation, oxygen removal and ultimately structural transition by elimination of the interlayer shear plane. Despite the process being commonly used, little is known about the microscopic mechanisms and hence factors that could ease the conversion. Here we aim to derive such insights via density functional theory calculations of the defect chemistry of the oxide and the oxynitride as well as the oxide's surface chemistry. Our results point to the crucial role of surface oxygen vacancies in forming clusters of \ce{NH3} decomposition products and in incorporating N, most favorably substitutionally at the anion site. N then spontaneously diffuses away from the surface, more easily parallel to the surface and in interlayer regions, while  diffusion perpendicular to the interlayer plane is somewhat slower. Once incorporation and diffusion lead to a local N concentration of about 70\% of the stoichiometric oxynitride composition, the nitridated oxide spontaneously transforms to a nitrogen-deficient oxynitride.
\end{abstract}

\maketitle

\section{Introduction}

Perovskite oxynitrides are promising photocatalysts, since their band structure is well-suited to both absorb solar light and drive the water-splitting redox reactions~\cite{EBBINGHAUS2009,Fuertes2012,Xie2013,Wang2020}. These materials can exist as nitrogen-rich \ce{ABN2O} or nitrogen-poor \ce{ABNO2} compositions, the latter generally being more active in photocatalytic applications~\cite{EBBINGHAUS2009}. While the A and B cations could, in principle, be selected from a large part of the periodic table~\cite{Castelli2012}, the compositions most popular in photocatalysis have \ce{Nb^5+}, \ce{Ta^5+} or \ce{Ti^4+} B sites, which are paired with A-site cations such as \ce{Sr^2+}, \ce{Ca^2+} or \ce{La^3+} respectively to respect charge neutrality. A particularly well-studied composition is \ce{LaTiO2N} (LTON)~\cite{EBBINGHAUS2009}.

\begin{figure}
	\centering
	\includegraphics[width=\columnwidth]{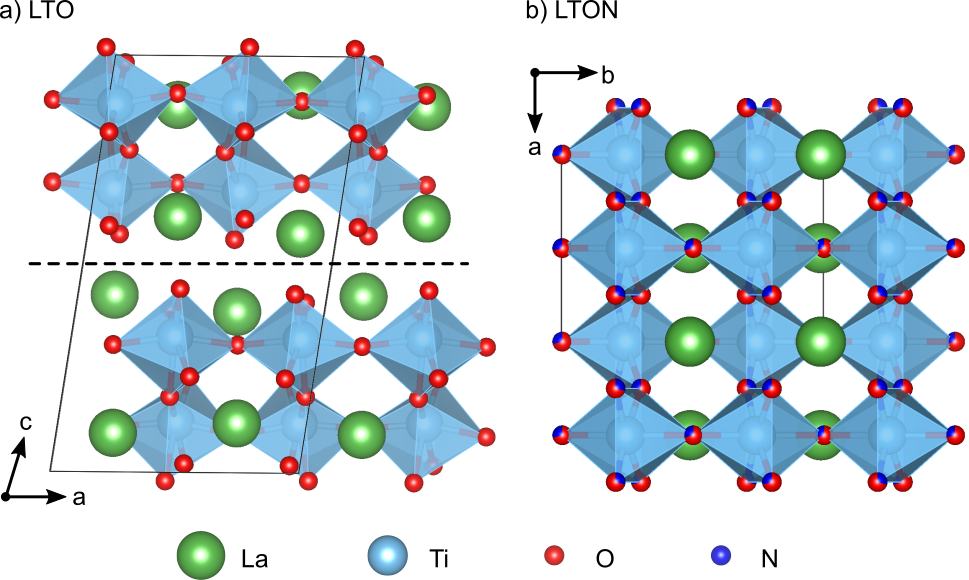}
	\caption{a) Monoclinic LTO unit cell with $P2_1$ space group. The dashed horizontal line indicates the interlayer plane parallel to (001). b) Orthorhombic LTON unit cell with $Imma$ space group and O/N disorder.}
	\label{fig:LTO_LTON}
\end{figure}

LTON is most often synthesized via thermal ammonolysis of a \ce{La2Ti2O7} (LTO) oxide precursor~\cite{EBBINGHAUS2008, EBBINGHAUS2009, Pokrant2014}. LTO belongs to the family of layered Carpy-Galy perovskite oxides~\cite{carpy1972contribution, carpy1973systeme, NUNEZVALDEZ2019181}, which are constituted of slabs of distorted corner-sharing \ce{TiO6} octahedra stacked along the $c$-axis (Fig.~\ref{fig:LTO_LTON}a), subsequent slabs being linked via La-O bonds across the interlayer plane~\cite{Ishizawa1982, Schmalle1993, Hwang2003}. These bonds are known to be weak, leading to preferential cleavage of microscopic LTO crystals along the interlayer plane~\cite{pokrant2016}. During ammonolysis LTO is placed under a flow of ammonia (\ce{NH3}) at temperatures higher than 600$^\circ$C for several hours. This process leads to formation of LTON, which has an orthorhombic perovskite structure with a mixture of O and N anions (Fig.~\ref{fig:LTO_LTON}b). While a \textit{cis} order is preferred within the individual octahedra~\cite{attfield2013principles}, LTON was shown to assume only partial~\cite{LOGVINOVICH2009, CHEN20162} or no long-range anion order~\cite{Clarke2002, Yashima2010}.

The transformation of LTO to LTON is a highly complex process, that has to involve the substitution of O by N, removal of excess O (the cation:anion ratio is 1:3.5 in LTO and 1:3 in LTON) and ultimately the corner-sharing linking of octahedra across the interlayer plane~\cite{EBBINGHAUS2008, Pokrant2014, pokrant2016}. At high temperature the \ce{2NH3 <=> N2 + 3H2} equilibrium  is shifted towards the products and \ce{NH2}, \ce{NH} and \ce{N} species may thus exist on LTO surfaces. Atomic and molecular hydrogen will form \ce{H2O} with lattice oxygen, creating oxygen vacancies. These vacancies are assumed to be instrumental for further \ce{NH3} decomposition and nitrogen incorporation, leading to a chain reaction. While initially any of the typically exposed LTO (100), (010) and (001) surfaces~\cite{pokrant2016} could interact with \ce{NH3} and its decomposition products, the transformation is associated with a volume shrinkage (see vertical dimensions in Figs.~\ref{fig:LTO_LTON}a and b) that leads to the formation of cracks that preferentially expose the LTO (001) interlayer plane~\cite{Pokrant2014}. A large proportion of N can thus be assumed to be incorporated by interaction of \ce{NH3} and its decomposition products with the (001) surface and subsequent diffusion into the layered LTO structure, while O atoms diffuse towards this surface and are removed as \ce{H2O}. This hypothesis is supported by the experimental observation that the oxynitride is not exclusively formed at the surface of the LTO particles, but stripes of LTON are observed extending into the bulk, following cracks oriented perpendicular to [001]~\cite{EBBINGHAUS2008}. Once these diffusion processes result in a composition around a interlayer plane sufficiently close to LTON, the octahedra are assumed to link via a relative lateral displacement of the adjacent slabs in the so-called zipper mechanism~\cite{EBBINGHAUS2008, Pokrant2014}.

Despite all experimental observations pointing towards the zipper mechanism, the indirect nature of the data does not yet allow a complete atomic-scale understanding of the transformation of LTO to LTON. From the above discussion, it is apparent that defects will play a crucial role in the \ce{NH3} decomposition, diffusion and zipper-mechanism aspects of the transformation mechanism. It is thus imperative to understand how O vacancies form in LTO, how they assist \ce{NH3} decomposition, how N is incorporated into LTO and how the anionic species diffuse in the layered structure. On the other hand, defects in LTON, resulting for example from incomplete transformation, also affect the final chemical and physical properties of the oxynitride. In particular non-stoichiometry or structural defects in LTON were shown to be detrimental for the photocatalytic performance~\cite{Pokrant2014, Lin2020}. Nevertheless, other studies report better photocatalytic performance for LTON containing less than the stoichiometric amount of nitrogen (about 70-80\%), despite the crystalline quality being inferior compared to stoichiometric LTON~\cite{Kasahara2003, MAEGLI2012}. 

In this work, we aim to gain insights on the transformation mechanism from LTO to LTON via density-functional theory (DFT) calculations. Our approach is to understand the defect chemistry of both the oxide precursor and the formed oxynitride as well as the diffusion kinetics of the most stable defect species. This is motivated by the fact that a highly defective LTO will eventually transform into an also defective LTON, our results providing an estimation of the required defect concentrations. We also study the interaction of \ce{NH3} and its decomposition products with the LTO (001) surface to establish the role of surface defects in N incorporation. A deeper knowledge of the defect chemistry and thermodynamics of these materials will pave the way towards a better understanding of the transformation mechanism and of the properties of the resulting oxynitride. These insights are key to identify strategies and guidelines for the synthesis design of oxynitrides with improved photo-electrochemical or photocatalytic activity.

\section{Methods}

Density functional theory (DFT) calculations were performed  with the \textsc{Quantum ESPRESSO} package~\cite{giannozzi2009quantum,Giannozzi2017} using the  PBE~\cite{PerdewPRL1996} exchange-correlation functional with a Hubbard \textit{U} correction~\cite{anisimov1991band, Dudarev1998} of 3 eV on the Ti 3\textit{d} states. Ultrasoft pseudopotentials~\cite{vanderbilt1990soft} with La(5\textit{s}, 5\textit{p}, 6\textit{s}, 5\textit{d}), Ti(3\textit{s}, 3\textit{p}, 4\textit{s}, 3\textit{d}), O(2\textit{s}, 2\textit{p}), and N(2\textit{s}, 2\textit{p}) valence states were employed, while wave functions and the augmented density were expanded in plane waves up to cutoffs of 40 and 320 Ry, respectively. 

\begin{figure}
	\centering
	\includegraphics[width=\columnwidth]{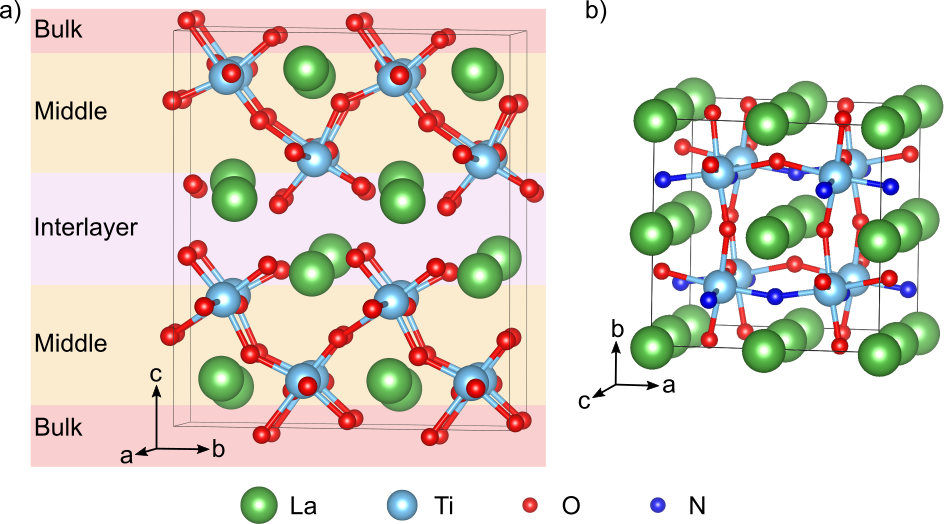}
	\caption{a) $1\times2\times1$ LTO supercell. Oxygen atoms colored in pink, orange, and red correspond to O sites lying in the interlayer, the middle and the bulk layer respectively. b) Pseudo-cubic LTON cell with \textit{cis} anion order.}
	\label{fig:LTO_LTON_cells}
\end{figure}

LTO was modeled using a $1\times2\times1$ supercell containing 88 atoms derived from the 44-atom unit cell (space group, $P2_1$)~\cite{Schmalle1993} (Fig.~\ref{fig:LTO_LTON_cells}a). A $3\times2\times3$ Monkhorst–Pack \textbf{k}-point mesh~\cite{monkhorst1976special} was used to sample the Brillouin zone of this supercell. Anion vacancy and substitutional defects in various charge states were created on the 14 symmetry inequivalent anion sites in this structure, while for anion interstitials we considered 22 different sites. To simplify the discussion, we will designate defects as lying either in the interlayer, the middle or bulk layer as shown in Fig.~\ref{fig:LTO_LTON_cells}a. Defect pairs of vacancies, substitution and interstitial defects were also considered.

LTON has an orthorhombic 20-atom unit cell with a longer $b$-axis~\cite{Clarke2002,Yashima2010} containing octahedral rotations and more importantly anion (dis)order (Fig.~\ref{fig:LTO_LTON}b). For our calculations we use a 40-atom $\sqrt{2}\times2\times\sqrt{2}$ pseudo-cubic supercell (Fig.~\ref{fig:LTO_LTON_cells}b) with a \textit{cis} anion order that was shown to be more favorable compared to a \textit{trans} anion arrangement~\cite{Ninova2017}. Reciprocal space was integrated using a $4\times4\times4$ Monkhorst–Pack k-point grid~\cite{monkhorst1976special}. For anion vacancy and substitutional defects we considered the two symmetry inequivalent sites, while for interstitials 16 different sites were sampled. In addition, we also considered cation antisite defects (\ce{La_{Ti}} and \ce{Ti_{La}}) in LTON along with pairs of the aforementioned defects.

For stoichiometric cells, both ionic positions and cell parameters were optimized, while for defective calculations, only atomic positions were relaxed with lattice vectors fixed to those of the respective stoichiometric cell. Structural relaxation was performed until the total energy and forces converged below $1.4\times10^{-5}$~eV  and $5\times10^{-2}$~eV/\AA, respectively.

We will refer to defects in a Kröger-Vink-like~\cite{KROGER1956307} notation ($X_s^q$), where the main letter ($X$) refers to the defect species, the subscript ($s$) to the defect site and - deviating from the formal notation - the numeric superscript ($q$) to the total charge of the simulation cell. The formation energy~\cite{freysoldt2014first} of such a defect is computed as
\begin{equation}
	\Delta E_{f,X_s^q} = E_{\mathrm{tot},X_s^q} - E_{\mathrm{tot,stoi}} -\sum_i n_i \mu_i + qE_\mathrm{F} + E_{\mathrm{corr}}\,
	\label{eq:formenerg}
\end{equation}
where $E_{\mathrm{tot},X_s^q}$ and $E_\mathrm{tot,stoi}$ are the DFT total energies of the defective and stoichiometric supercells, respectively. $E_\mathrm{F}$ is the Fermi energy relative to the valence band maximum of the stoichiometric  system, which can assume values within the band gap $E_\mathrm{g}$ ($0 \leq E_\mathrm{F} \leq E_\mathrm{g}$) of the non-defective structure. $E_\mathrm{corr}$ is a corrective term necessary for charged defects (with an automatically added neutralizing background) to align the electrostatic potential of the defective charged cell with the one of the neutral stoichiometric cell. It was obtained by calculating the electrostatic potential difference between the two systems via averaging the electrostatic potential in spheres around atomic sites located far from the defect~\cite{lany2008}. No further finite-size corrections~\cite{freysoldt2014first} were applied, given the relatively high dielectric constant of LTO (42-62~\cite{Hojamberdiev2015}) and LTON (750~\cite{Lin2020}). Finally, $n_i$ indicates the number of atoms of a certain species $i$ that are added ($n_i > 0$) or removed ($n_i < 0$) from the supercell to form the defect and $\mu_i$ the species' chemical potential. We define the chemical potential as $\mu_i = \mu_i^0 + \Delta\mu_i$, where $\mu_i^0$ is the chemical potential in a reference state and $\Delta\mu_i$ varies within thermodynamic stability limits and directly relates to the chemical environment. Pure metals are used as reference state for La and Ti, the \ce{O2} molecule for O and \ce{NH3} (LTON is synthesized under ammonia flow) or \ce{N2} (considered as the nitridating agent) for N. The \ce{NH3} reference yields an N chemical potential 1.01 eV higher than the \ce{N2} reference. We restrict the various $\Delta\mu_i$ to regions of the phase diagrams where LTO/LTON is stable and where the formation of competing phases, such \ce{La2O3} and \ce{TiO2}, is not favorable (ESI\dag\ Section~\ref{sec:phasediagram}). Considering typical conditions during thermal ammonolysis~\cite{Pokrant2014,pokrant2016} and the typical n-type nature of transition metal oxides/oxynitrides, we will, in the main text, show results in the N-rich limit and for a Fermi energy at the conduction band edge. More detailed results as a function of these parameters can be found in the ESI\dag\ Sections~\ref{sec:SI_LTO} and~\ref{sec:SI_LTON}.

Diffusion barriers were calculated using the climbing-image nudged elastic band (NEB) method~\cite{Henkelman2000}. The number of images was selected for an initial inter-image distance of about 0.25 \AA\ and the path was optimized until the forces on each images converged to $1\times10^{-3}$~eV/\AA.

The LTO (001) surface was modeled as a one LTO-layer thick slab with the bottom half of the atoms fixed at bulk positions, a 10 \AA\ vacuum gap and a dipole correction~\cite{bengtsson1999dipolecorr}. Reciprocal space of this 44-atom cell was sampled with a $3\times4\times1$ k-point mesh~\cite{monkhorst1976special}. Adsorption energies were calculated as
\begin{equation}
	\Delta E_\mathrm{ads} = E_\mathrm{slab+ads} - E_\mathrm{ads} - E_\mathrm{slab},
\end{equation}
where $E_\mathrm{slab+ads}$, $E_\mathrm{ads}$ and $E_\mathrm{slab}$ refer to the DFT total energies of the slab with adsorbate, the isolated adsorbate and the clean slab respectively. \textit{Ab-initio} atomistic thermodynamics~\cite{reuter_first-principles_2003, reuter2005ab} as implemented in the Atomistic Simulation Environment (ASE)~\cite{ase-paper} were used to account for temperature and partial pressure effects on adsorption.

The optical properties of defective LTON were calculated based on the frequency dependent dielectric matrix within the VASP package~\cite{Kresse:1993ty, Kresse:1994us, Kresse:1996vk, Kresse:1996vf}, based the PBE~\cite{Perdew:2008fa} exchange-correlation functional using PAW potentials~\cite{Blochl:1994uk, Kresse:1999wc} with La(5\textit{s}, 5\textit{p}, 5\textit{d}, 6\textit{s}), Ti(3\textit{s}, 3\textit{p}, 3\textit{d}, 4\textit{s}), O(2\textit{s}, 2\textit{p}) and N(2\textit{s}, 2\textit{p}) valence shells together with a plane-wave cutoff of 500 eV. A rotationally invariant~\cite{Dudarev1998} Hubbard \textit{U} correction~\cite{anisimov1991band} of 3~eV was applied to the Ti 3\textit{d} orbitals. All calculations were performed with a spin-polarised setup using structures previously relaxed in \textsc{Quantum ESPRESSO}. The number of bands was about tripled (600) from the default value (216). Absorption spectra were extracted using the vaspkit package~\cite{vaspkit2021}.

LTON powders are synthesized by thermal ammonolysis of an oxide precursor, LTO. The precursor oxide is produced by a solid-state approach where 12.5 mmol \ce{TiO2} (Anatase, Sigma Aldrich, 99.9\%), 6.25 mmol \ce{La2O3} (Sigma Aldrich, 99.9\%), and 12.5 mol \ce{NaCl} (VWR, 99\%) are mechanically mixed using a roll-mill and calcined for 10 h at 1200 $^\circ$C. After calcination the flux is removed and the resulting \ce{La2Ti2O7} is dried at 100 $^\circ$C for 12 h. For thermal ammonolysis, 1 g of \ce{La2Ti2O7} is placed inside an alumina tube which is purged with \ce{N2} for 30 min and \ce{NH3} for 40 min at a flow rate of 0.2 L/min. The synthesis conditions are varied by applying various temperatures and durations (1000 $^\circ$C for 16 h, 18 h, and 25 h, 1050 $^\circ$C for 18 h and 25 h) while keeping a constant \ce{NH3} flow of 0.2 L/min. UV-vis diffuse reflectance spectra are collected using an UV-vis-NIR spectrophotometer (Perkin Elmer, Lambda 1050) equipped with an integrating sphere over a spectral range of 200-900 nm (step size 2 nm). \ce{BaSO4} is used as a reference. The spectra are transformed to the Kubelka-Munk function~\cite{weckhuysen1999recent} and normalized before plotting.

All data are available on the Materials Cloud archive~\cite{archive}.

\section{Results and Discussion}

\subsection{Interaction of LTO (001) with \ce{NH3}}

The (001) shear interface is known to be the preferred cleaving plane for LTO and can hence be assumed as the dominant surface interacting with \ce{NH3} and its decomposition products during ammonolysis. The (001) surface can either be the actual surface of a particle or exposed in cracks formed during ammonolysis~\cite{pokrant2016}. In order to gain insights into the interaction and decomposition of \ce{NH3} in contact with this surface and the role oxygen vacancies play in this process, we study the interaction of ammonia and its dehydrogenated derivatives with pristine and oxygen deficient LTO (001) surfaces. We find the energetically most favorable oxygen vacancy to reside at the very surface (ESI\dag\ Table~\ref{tbl:surf_VO_form} and Fig.~\ref{fig:surf_VO}). Since in LTO without a surface (Section~\ref{sec:defLTO}) sites away from the interlayer plane are preferred, this preference likely stems from the reduced number of broken bonds at the surface.

We find molecular \ce{NH3} to adsorb most favorably at a Ti-top site on the stoichiometric surface with an adsorption energy of 0.71 eV, while on the defective surface it adsorbs most favorably above the \ce{V_O} with an adsorption energy of 1.75 eV (Table~\ref{tbl:molecular_ads_energy}). On the defective surface, an \ce{NH3} adsorbed at the \ce{Ti_1}-top site spontaneously migrates into the \ce{V_O} as evident by the almost equivalent adsorption energy. This clearly indicates the importance of \ce{V_O} for the adsorption of ammonia as once adsorbed, transition state theory predicts the molecule to remain bound on the microsecond time scale even at elevated ammonolysis temperatures, whereas residence times in the sub-nanosecond range result on the stoichiometric surface. This is also the case for the decomposition products \ce{NH2}, \ce{NH} and \ce{N} that, when adsorbed adjacent to a vacancy, spontaneously relax into the vacancy, clearly demonstrating this to be the most favorable adsorption site.

\begin{table}
	\centering
	\caption{Adsorption energies (in eV) of \ce{NH3} on the stoichiometric and defective LTO (001) surface for the sites presented in ESI\dag\ Fig. \ref{fig:adsorption_sites}. The most relevant sites are highlighted in bold.}
	\label{tbl:molecular_ads_energy}
	\begin{tabular}{l|rr}
		\hline
		\hline
		Site          & Stoichiometric  & Defective \\
	  	\hline 
		\ce{Ti_1}-top & \textbf{-0.71} & \textbf{-1.75} \\
		\ce{Ti_2}-top & \textbf{-0.71} & -0.66 \\
		\ce{La_1}-top & -0.53 & -1.38 \\
		\ce{La_2}-top & -0.63 & -1.26 \\
		Ti-La-bridge  & -0.24 & -1.26 \\
		Ti-Ti-bridge  & -0.65 & -0.99 \\
		La-La-bridge  & -0.67 & -1.01 \\
		\ce{V_O}      & -     & \textbf{-1.76} \\
		\hline
	\end{tabular}
\end{table}

\begin{figure}
\centering
\includegraphics[width=\columnwidth]{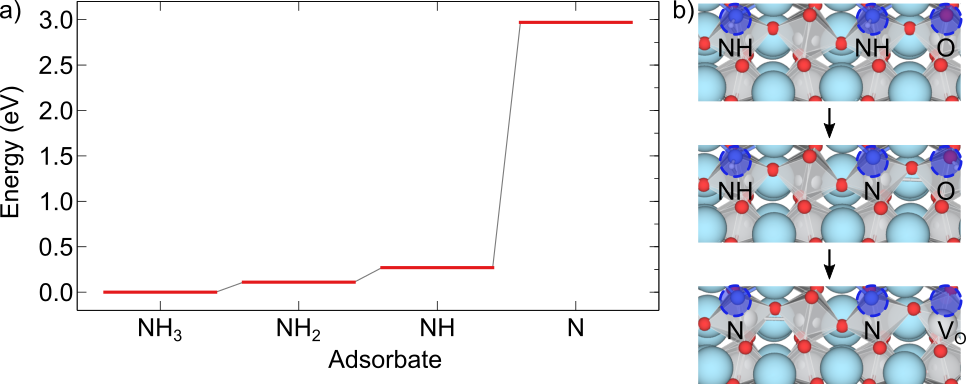}
\caption{a) Energy profile for the decomposition reaction of \ce{NH3} to \ce{N} and b) alternative mechanism with two nearby \ce{NH} for the final step in the decomposition reaction. The relevant sites are labeled and highlighted by blue circles.}
\label{fig:energy_profile}
\end{figure}

Next, we calculated the thermodynamics of the decomposition reaction of \ce{NH3} at a surface \ce{V_O}. We initially consider that for each decomposition step half an oxygen vacancy forms and half a water molecule is released as shown by reactions~\ref{eq:rx1} to~\ref{eq:rx3}.
\begin{alignat}{5}
		\frac{3}{2} \mathrm{stoi} + \mathrm{def_{\ce{NH3}}}
	&\rightarrow             &\mathrm{stoi} + &&\mathrm{def_{\ce{NH2}}} + &&\frac{1}{2} \mathrm{def} + &&\frac{1}{2} \ce{H2O} \label{eq:rx1}\\
	&\rightarrow \frac{1}{2} &\mathrm{stoi} + &&\mathrm{def_{\ce{NH}}}  + &&            \mathrm{def} + &&            \ce{H2O} \label{eq:rx2}\\
	&\rightarrow             &                &&\mathrm{def_{\ce{N}}}   + &&\frac{3}{2} \mathrm{def} + &&\frac{3}{2} \ce{H2O} \label{eq:rx3}
\end{alignat}
Here $\mathrm{stoi}$ stands for the stoichiometric slab, $\mathrm{def}$ for a slab with an oxygen vacancy and $\mathrm{def_{prod}}$ for a slab with the respective decomposition product (prod = \ce{NH3}, \ce{NH2}, \ce{NH}, \ce{N}) adsorbed at the vacancy site. We compute the energy profile for the reaction, considering temperature and partial pressure effects via \textit{ab-initio} atomistic thermodynamics by setting $T = 1223$ K, $p_{\ce{NH3}} = 10^5$ Pa and $p_{\ce{H2O}} = 10$ Pa, which are conditions relevant for ammonolysis. The resulting energy profile in Fig.~\ref{fig:energy_profile}a shows that while most of the decomposition steps have a moderate energy cost, the final conversion of \ce{NH} to \ce{N} is prohibitively large with a height in excess of 2 eV, likely due to the formation of a strongly undercoordinated N atom.

The above reactions, however, also show that the initial facile decomposition steps lead to formation of additional \ce{V_O} adjacent to the initial \ce{NH3} adsorbate. Clusters of \ce{V_O} that will partially be filled with additional decomposition products are thus likely to form on the LTO surface. It makes, therefore, sense to consider if the final step could be facilitated in presence of two \ce{NH} adsorbed in nearby \ce{V_O}. The reaction we study proceeds via transfer of one H from N to an adjacent O followed by the transfer of the second H to the same O, desorption of \ce{H2O} and formation of the oxygen vacancy. This process is shown in Fig.~\ref{fig:energy_profile}b and has a thermodynamic cost of 1.12 eV, only half of the one in Fig.~\ref{fig:energy_profile}a.

These results highlight that \ce{V_O} are crucial to adsorb \ce{NH3} and decompose it to \ce{NH}, forming additional \ce{V_O} in the process. The proximity of these newly formed \ce{V_O} is then crucial to facilitate the final decomposition step and incorporation of N into LTO. This model shows that surface \ce{V_O} will nucleate regions in which N is incorporated in a chain-like reaction, in agreement with the experimental observation that large regions of LTO simultaneously transform to LTON~\cite{pokrant2016}.

\subsection{Defect chemistry of LTO}\label{sec:defLTO}

\begin{figure}
	\centering
	\includegraphics{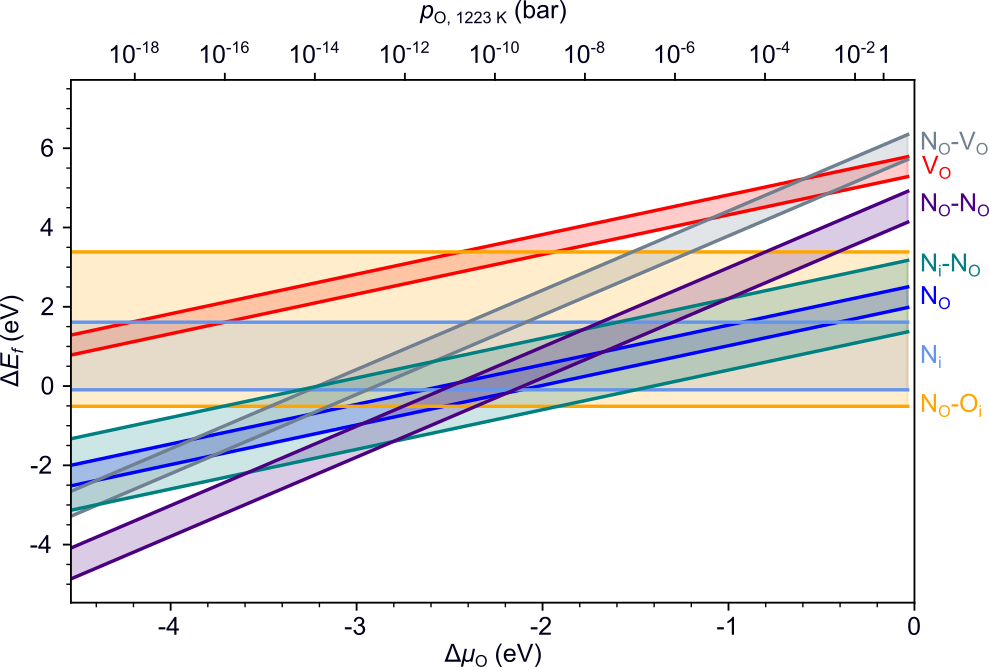}
	\caption{Formation energies of various defects and defect pairs in LTO as a function of the oxygen chemical potential and for a Fermi energy at the conduction band minimum. The spread in values shown by the highlighted area between the respective minimum and maximum formation energy is due to the different anion sites in the layered structure.}
	\label{fig:LTO_summary}
\end{figure}

Figure \ref{fig:LTO_summary} summarizes the formation energies of all investigated defects and defect pairs in LTO as a function of the oxygen chemical potential and evaluated for experimentally relevant n-type conductivity conditions with the Fermi energy at the conduction band minimum. Details for each defect can be found in the ESI\dag\ Section~\ref{sec:SI_LTO} with specific references given below.

\textbf{\ce{V_O}}: For experimentally relevant conditions, we predict oxygen vacancies (\ce{V_O}) to be in their neutral charge state (ESI\dag\ Fig.~\ref{fig:Ef_VO_LTO}a). \ce{V_O} in this charge state are most stable in the middle layer  and least stable at the interlayer plane (ESI\dag\ Fig.~\ref{fig:Ef_VO_LTO_confs}a). Even under O-poor conditions (towards the left in Fig.~\ref{fig:LTO_summary}, \ce{V_O} have a relatively high formation energy between $\sim$0.8 eV in the center of the slab and $\sim$1.3 eV at the interlayer plane (Table~\ref{tbl:LTO_VOconfs}), suggesting a rather low concentration of oxygen vacancies in LTO. This lack of \ce{V_O}, in particular at the interlayer plane, will hinder the zipper mechanism that relies on this defect. Our data also implies that surface \ce{V_O} need to be created by \ce{NH3} decomposition at the surface rather than by creation within LTO and subsequent diffusion to the surface.

\textbf{\ce{N_O}, \ce{N_i} and \ce{N_O-O_i}}: We studied nitrogen substitution at an O site (\ce{N_O}) and interstitial nitrogen (\ce{N_i}) as two possible ways to insert N into LTO. For experimentally relevant conditions, we predict the \ce{N_O^{-1}} charge state to be most stable (ESI\dag\ Fig.~\ref{fig:Ef_NO_LTO}), which corresponds to an ionic \ce{N^3-} on the anion site. This defect spontaneously forms already for intermediate oxygen chemical potentials. For interstitial N we also predict a fully ionic \ce{N_i^{-3}} charge state to be most stable (ESI\dag\ Fig.~\ref{fig:Ef_Ni_LTO}a). For this defect, we observe strong structural rearrangements, the N atom displacing an O from its lattice site to an interstitial site. This is in line with XPS results for N-doped LTO that indicate the environment of La ions in the structure not to be affected by doping and that N are bound to Ti atoms in the Ti octahedra~\cite{FUJISHIMA2008, meng2012}. The \ce{N_i} defect should thus correctly be labeled as an \ce{(N_O-O_i)^{-3}} defect pair. The relative stability of these two N-related defects depends on the experimental conditions (Fig.~\ref{fig:LTO_summary}): \ce{N_O^{-1}} are more likely to be observed under O-poor conditions, while \ce{N_i}/\ce{(N_O-O_i)^{-3}} are more stable in O-rich environments. In both cases, the N atom is preferentially found in the middle-layer (ESI\dag\ Figs.~\ref{fig:Ef_NO_LTO_confs}c and~\ref{fig:LTO_Niconfs}c), suggesting that the nitridating \ce{N^3-} species can diffuse away from the LTO interlayer interface, where, as we will show below, its mobility is highest. We investigated additional configurations for \ce{N_O-O_i} (shown in orange in Fig.~\ref{fig:LTO_summary}), confirming the -3 charge state while finding somewhat lower formation energies (ESI\dag\ Section~\ref{sec:NOOi}) for configurations where \ce{N_O} and \ce{O_i} can form a bond. This limits the isolated substitutional N to fairly O-poor conditions and confirms the strong tendency of N to replace O, leading to mobile interstitial O species and anion superstoichiometry prior to LTON formation.

\textbf{\ce{N_O-V_O}, \ce{N_i-N_O} and \ce{N_O-N_O}}: The above results highlight the potential importance of defect pairs for nitrogen incorporation. We therefore also investigated substitutional nitrogen paired up with either an oxygen vacancy, a nitrogen interstitial or another substitutional nitrogen. In all cases the most stable charge state corresponds to fully ionic \ce{N^3-} under experimentally relevant conditions (ESI\dag\ Figs.~\ref{fig:Ef_NOVO_LTO}, \ref{fig:Ef_NiNO_LTO} and \ref{fig:Ef_NONO_LTO}). Our results (Fig.~\ref{fig:LTO_summary}) show that at very low oxygen chemical potential, \ce{V_O} can lower the formation energy of a \ce{N_O} compared to isolated \ce{N_O}. Intrinsic LTO defects may hence thermodynamically assist nitrogen incorporation. A \ce{N_i} paired with a \ce{N_O} may also lead to a lower formation energy than isolated \ce{N_O} (Fig.~\ref{fig:LTO_summary}), in particular also in the boundary layer, if the two defects are close (ESI\dag\ Fig.~\ref{fig:Ef_NiNO_LTO_confs}d). This implies that nitrogen incorporation in form of interstitials may be assisted by already existing substitutional nitrogen. Finally, we find that two substitutional N can coexist in the middle layer, however, most favorably so at large separation  (ESI\dag\ Fig.~\ref{fig:Ef_NONO_LTO_confs}b). At very low oxygen chemical potential this is preferred over isolated \ce{N_O}.

\textbf{Summary:} Our results for LTO show that the oxygen vacancy concentration is fairly low, in particular around the interlayer plane. Nitrogen is most favorably incorporated by substituting an oxygen under the low oxygen partial pressure situation of ammonolysis, the displaced oxygen becoming a mobile interstitial species. Once in the LTO lattice, substitutional N will not tend to cluster but may ease interstitial N incorporation in their vicinity. 

\subsection{Nitrogen diffusion in LTO}

\begin{table}
	\centering
	\caption{Range of diffusion barriers for diffusion either parallel or perpendicular to the interlayer plane and in/between given LTO layers. See ESI\dag\ Fig.~\ref{fig:NEB} and Table~\ref{tbl:NEB} for full data.}
	\label{tbl:diffusion}
	\begin{tabular}{l|r}
		\hline
		\hline
		Path                                      & Barrier range (eV) \\
	  	\hline 
		Parallel, interlayer                      & 0.01-0.79 \\
		Parallel, middle layer                    & 0.50-1.99 \\
		Parallel, bulk layer                      & 0.78-1.09 \\
		Perpendicular, interlayer to middle layer & 0.77-1.67 \\
		Perpendicular, within middle layer        & 0.77-1.67 \\
		Perpendicular, middle to bulk layer       & 1.32-1.96 \\
		\hline
	\end{tabular}
\end{table}

Once nitrogen enters LTO as a substitutional N at a (001) crack face or surface, it needs to diffuse deeper into the LTO layered structure to induce the conversion to the oxynitride. We study diffusion of \ce{N_O} via a \ce{V_O} diffusion vehicle using the NEB method, considering different arrangements of these two defects as well as different charge states. In order to estimate facile diffusion directions, we then use these elemental diffusion events to build migration paths parallel to the interlayer plane in the interlayer, middle and bulk layer as well as paths perpendicular to the interlayer plane. The range of barriers involved in these paths are shown in Table~\ref{tbl:diffusion}.

The lowest migration barriers are observed for diffusion along the interlayer plane, further suggesting these layers to be rapid diffusion directions, which is in line with elongated pores along this direction and the fact that the LTO to LTON transformation occurs also far from the surface~\cite{EBBINGHAUS2008}. Diffusion parallel to the interlayer plane then becomes increasingly more difficult the further \ce{N_O} is from the interlayer plane. Diffusion barriers perpendicular to the interlayer plane are similar to the parallel ones in deeper layers. This data suggests that  while \ce{N_O} is more stable away from the interlayer plane, diffusion to these sites is rather slow. An N-saturated interlayer plane and thus higher driving force, as well as high ammonolysis temperatures are thus prerequisite for N-diffusion away from the interlayer plane.

\subsection{Defect chemistry of LTON}

\begin{figure}
	\centering
	\includegraphics{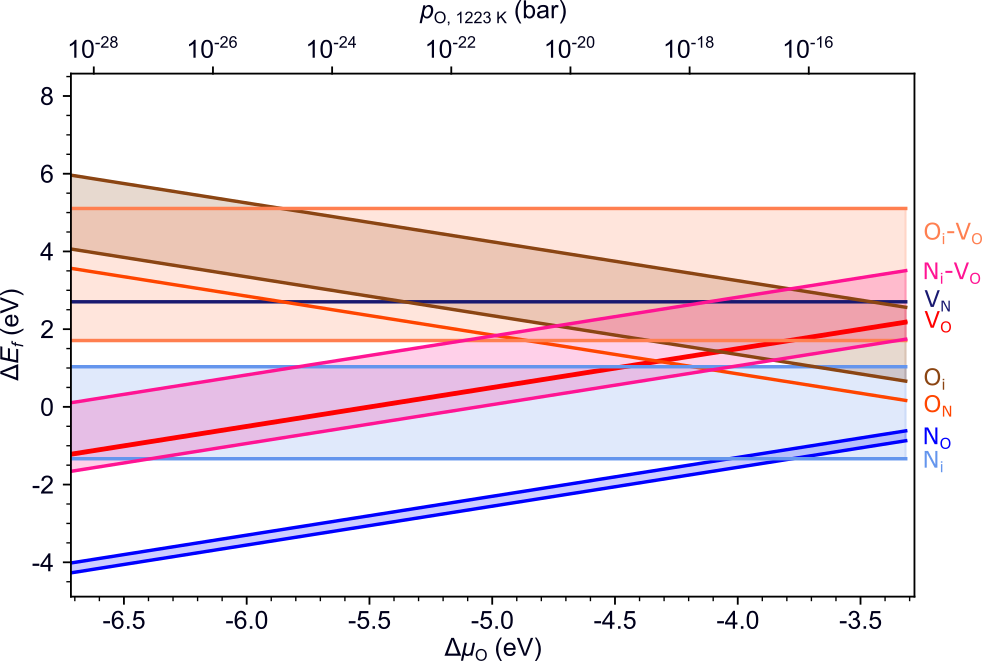}
	\caption{Formation energies of various defects and defect pairs in LTON as a function of the oxygen chemical potential and for a Fermi energy at the conduction band minimum. The spread in values shown by the highlighted area between the respective minimum and maximum formation energy stems from the different anion sites in the structure.}
	\label{fig:LTON_summary}
\end{figure}

Figure \ref{fig:LTON_summary} summarizes the formation energies of all anionic defects and defect pairs in LTON as a function of the oxygen chemical potential and evaluated for experimentally relevant n-type conductivity conditions with the Fermi energy at the conduction band minimum. Details for each defect, can be found in ESI\dag\ Section~\ref{sec:SI_LTON} with specific references given below. Also shown in ESI\dag\ Sections~\ref{sec:antisites1}~and \ref{sec:antisites2} is data for cationic antisite defects, as these defects, while potentially promoted by pairing with anion substitutions, have rather high formation energies and are thus not relevant during ammonolysis.

\textbf{\ce{V_O} and \ce{V_N}}: The formation energy for a neutral \ce{V_O}, relevant under experimental conditions (ESI\dag\ Fig.~\ref{fig:Ef_VO_LTON}), is very similar on the two inequivalent sites as shown by the narrow range in Fig.~\ref{fig:LTON_summary}. Comparing the formation energy in LTON to the one in LTO (Fig.~\ref{fig:LTO_summary}) under similar conditions, the \ce{V_O} formation energy is about 1~eV higher in the oxynitride. The \ce{V_O} is thus preferentially located in LTO, where it can be annihilated during N incorporation. As expected under N-rich ammonolysis conditions, \ce{V_O} formation is favored by 1-2~eV compared to \ce{V_N} formation, the \ce{V_N} being in a -1 charge state under these conditions (ESI\dag\ Fig.~\ref{fig:Ef_VN_LTON}). Not only is this important for the transformation of LTO to LTON, but the formation of \ce{V_N} would also lead to in-gap states (ESI\dag\ Fig.~\ref{fig:PDOS_VO_LTON}) which are possibly detrimental for the visible light absorption and hence the photocatalytic performance of the synthesized LTON.

\textbf{\ce{O_N} and \ce{N_O}}: Substitutional \ce{N_O} in the most favorable -1 charge state (ESI\dag\ Fig. \ref{fig:Ef_NO_LTON}a) are the most stable defects for a large range of the oxygen chemical potential, which is due to the considered N-rich ammonolysis environment. This defect will not induce in-gap states and not affect light absorption (ESI\dag\ Fig.~\ref{fig:PDOS_NO_LTON}a). Substitutional \ce{O_N} in its most favorable -1 charge state (ESI\dag\ Fig.~\ref{fig:Ef_ON_LTON}) is unlikely to form, unless under very O-rich conditions and would lead to doping into the conduction band (ESI\dag\ Fig.~\ref{fig:PDOS_ON_LTON}). 

\textbf{\ce{N_i}, \ce{O_i}, \ce{N_i-V_O} and \ce{O_i-V_O}}: N could also be incorporated in LTON as an \ce{N_i} interstitial. Due to the N-rich ammonolysis environment, \ce{N_i} can easily form in LTON (ESI\dag\ Fig.~\ref{fig:Ef_Ni_LTON}), especially as the nitridating \ce{N_i^{-3}} species, given sufficiently long ammonolysis times. \ce{N_i} prefers sites where it can form \ce{N2} dimers (ESI\dag\ Fig.~\ref{fig:Ef_Ni_LTON_confs}) and will lead to in-gap states (ESI\dag\ Fig.~\ref{fig:PDOS_Ni_LTON}) that are detrimental to the optical properties of LTON. \ce{N_i} formation could be further favored in presence of \ce{V_O} under O-poor conditions (ESI\dag\ Figs.~\ref{fig:Ef_Ni_LTON} and~\ref{fig:Ef_VONi_LTON}). However, even if \ce{N_i} could be easily formed, they would eventually convert to \ce{N_O} defects that are generally more stable (Fig.~\ref{fig:LTON_summary}). \ce{O_i} formation is less relevant, with fairly high formation energies for the most favorable \ce{O_i^{-2}} charge state (ESI\dag\ Fig.~\ref{fig:Ef_Oi_LTON}) except under very O-rich conditions. Like nitrogen, also \ce{O_i} prefers to form \ce{O2} dimers (ESI\dag\ Fig.~\ref{fig:Ef_Oi_LTON_confs}). Under O-poor conditions, the formation of \ce{O_i-V_O} Frenkel pairs becomes favorable, resulting in doping into the conduction band (ESI\dag\ Fig.~\ref{fig:PDOS_VoOi_LTON}).

\textbf{Summary}: Our results show that substitutional \ce{N_O} is the most likely defect in LTON, leading to nitrogen superstoichiometry without affecting light absorption. Nitrogen interstitials could also form and will do so most favorable in the form of \ce{N2} dimers, inducing gap states detrimental to optical absorption. Nevertheless in presence of \ce{V_O} these defects are expected to eventually relax to \ce{N_O}.

\begin{figure}
	\centering
	\includegraphics{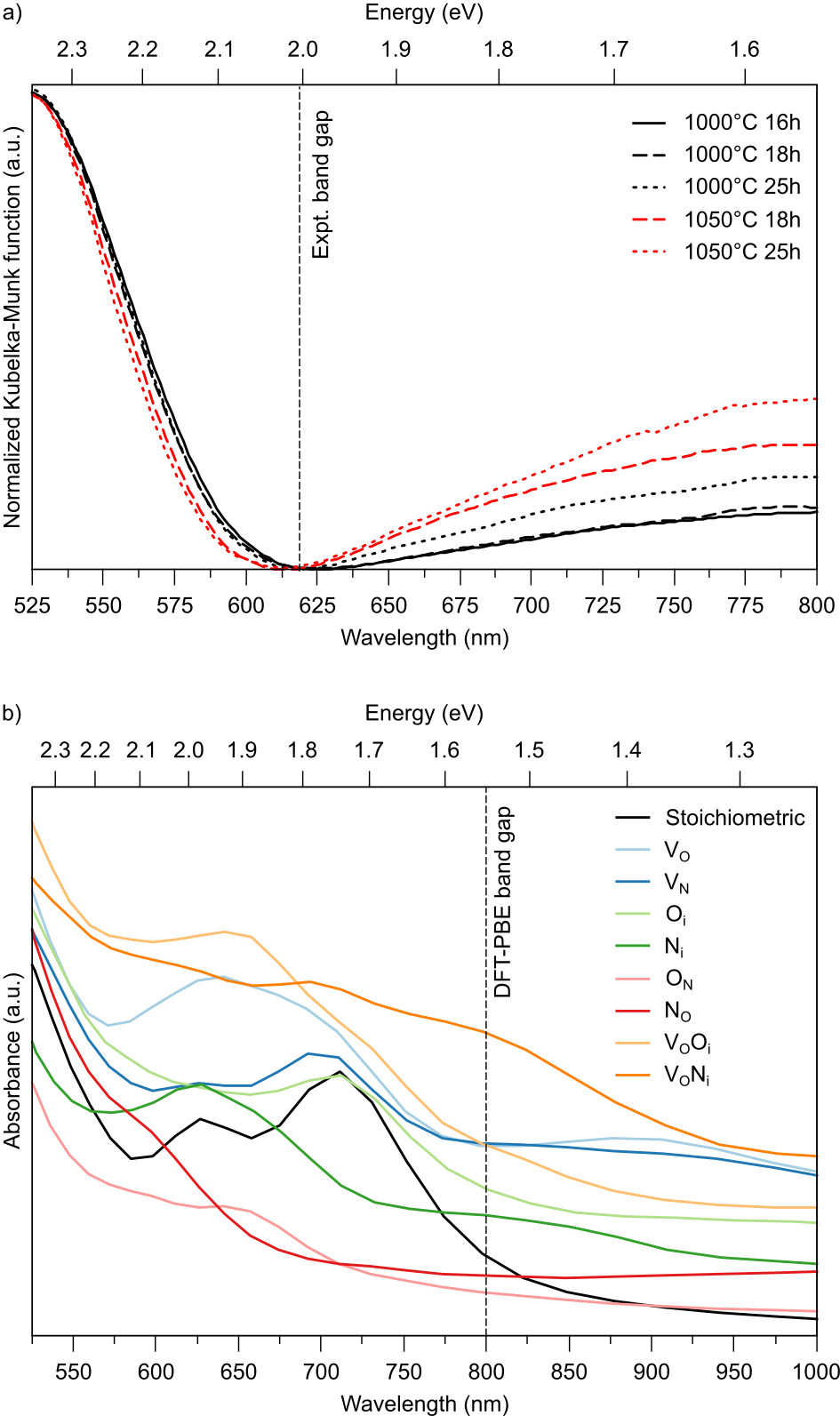}
	\caption{a) Experimental UV-vis spectra for LTON after different ammonolysis times and temperatures. b) Computed optical absorption spectra for stoichiometric and defective LTON.}
	\label{fig:expt}
\end{figure}

To support these computational results, we performed experiments with different ammonolysis times and temperatures. These show that both longer times or higher temperatures lead to emergence of an additional peak around 800 nm (1.55 eV) in the UV-vis spectra of LTON (Fig.~\ref{fig:expt}a), which likely stems from kinetically slow to form defects. To test this hypothesis, we computed the optical absorption spectra for pristine LTON as well as structures with the various relevant defects and defect pairs (Fig.~\ref{fig:expt}b). Even though a direct quantitative comparison with experiment is impossible due to the significantly underestimated band gap in our semi-local DFT calculations, we can see that anionic defects lead to absorption at lower energy (higher wavelength) compared to the pristine sample. While this effect is very minor for substitutional \ce{N_O} and \ce{O_N} defects, it is more pronounced for interstitial \ce{N_i} and \ce{O_i} defects. Vacancies (\ce{V_O} and \ce{V_N}) lead to even more significant contributions, that are highest when pairing with an interstitial occurs. Given the rather low bulk \ce{V_O} concentration, it seems reasonable to assign the experimental observation primarily to \ce{N_i} defects in metastable dimer configurations that have not yet relaxed to \ce{N_O}. This points to the interpretation that extended ammonolysis times or higher temperatures promote \ce{N_i} and \textit{local} N superstoichiometry, due to continuous N incorporation at the (crack) surface.

\subsection{Crossover defect concentration}
%
\begin{figure}
	\centering
	\includegraphics[width=\columnwidth]{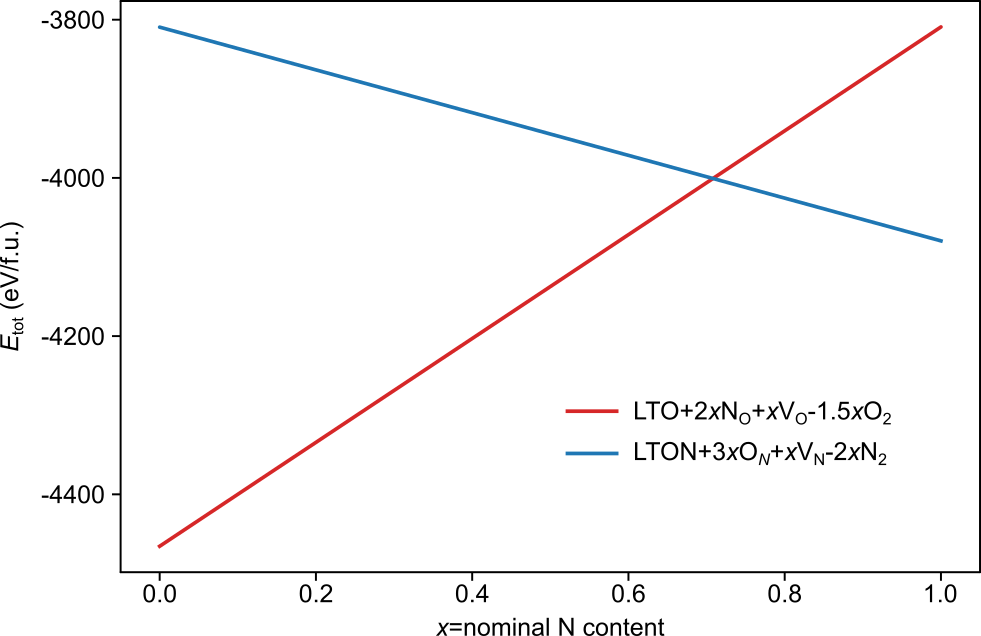}
	\caption{Total energy per formula unit of defective LTO (in red) and LTON (blue) cells obtained according to Eqs.~\ref{eq:crossover1} for LTO and ~\ref{eq:crossover2} for LTON, as a function of the nominal N content (x).}
	\label{fig:crossover1}
\end{figure}
The above data on the defect chemistry of both LTO and LTON can be used to estimate the crossover defect concentration, \textit{i.e.} the N content that thermodynamically stabilizes defective LTON compared to defective LTO. We consider the following two reactions (in proper Kröger-Vink notation) that alter the N content of LTO and LTON respectively, based on the most relevant substitutional \ce{N_O} and \ce{O_N} defects determined above:
\begin{align}
7\mathrm{O_X^X} + \mathrm{N_2} &\rightarrow 4 \mathrm{O_O^X} + 2 \mathrm{N_O'} + \mathrm{V_O^{\bullet \bullet}} + 1.5 \mathrm{O_2}\label{eq:crossover1}\\
4\mathrm{N_N^X} + 1.5 \mathrm{O_2} &\rightarrow 3 \mathrm{O_N'} + \mathrm{V_N^{\bullet\bullet\bullet}} + 2 \mathrm{N_2}\label{eq:crossover2}
\end{align}
We note that despite these reactions being written with \ce{N2} and \ce{O2} as molecular reactants/products our results are the same if \ce{NH3} and \ce{H2O} would be considered. This is achieved by estimating the total energy of a defective cell by adding the formation energy of the created defects to the total energy of the respective stoichiometric cell. Since these formation energies are calculated for relevant ammonolysis conditions, the appropriate chemical potentials of \ce{O} and \ce{N} containing molecular reactants/products are implicitly taken into account. We further note that this simple approach neglects defect-defect interactions that will become relevant for high defect concentrations, which we address below. Using this approach, Fig.~\ref{fig:crossover1} shows that defective LTON becomes more stable than defective LTO at about 70\% N substitution in LTO.

To further prove this finding and to include defect-defect interactions, we altered the $1\times2\times1$ LTO cell (88 atoms) to result in about 70\% N content by introducing 12 substitutional N and 6 oxygen vacancies. \ce{N_O} and \ce{V_O} were inserted in the interlayer (I), the middle (M) or the bulk (B) layer, resulting in either \textit{cis} or \textit{trans} local anion order as well as in a random fashion. For each defective LTO configuration, a corresponding $2\times2\sqrt{2}\times2\sqrt{2}$ LTON cell with the same number and types of atom was created in order to compare their total energies. 
\begin{figure}
	\centering
	\includegraphics[width=\columnwidth]{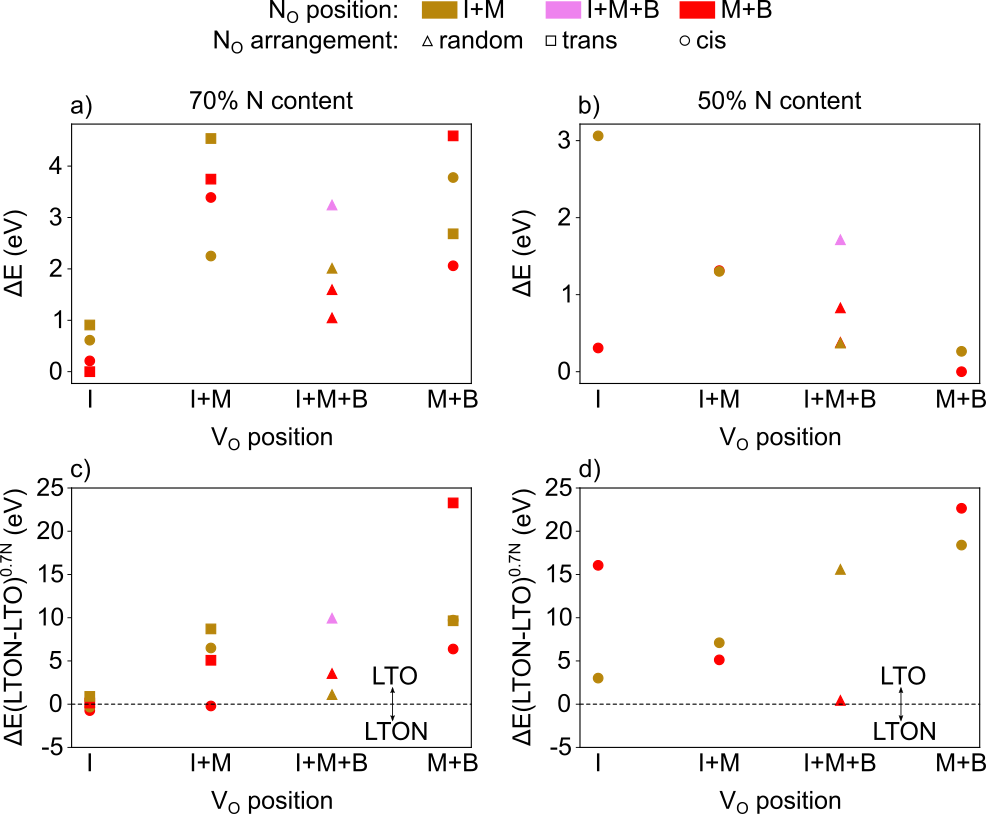}
	\caption{Energy of LTO cells with a) 70\% or b) 50\% nominal N content obtained by introducing \ce{N_O} and \ce{V_O} defects in the interlayer (I), the middle (M) and bulk (B) layer of LTO relative to the most stable cell for each N content. \ce{N_O} were arranged either in \textit{cis}, \textit{trans} or random fashion. Energy difference between an LTON and LTO cell with the same number and type of atoms for c) 70\% and d) 50\% nominal N content.}
	\label{fig:crossover2}
\end{figure}
\begin{figure*}
	\centering
	\includegraphics[width=\textwidth]{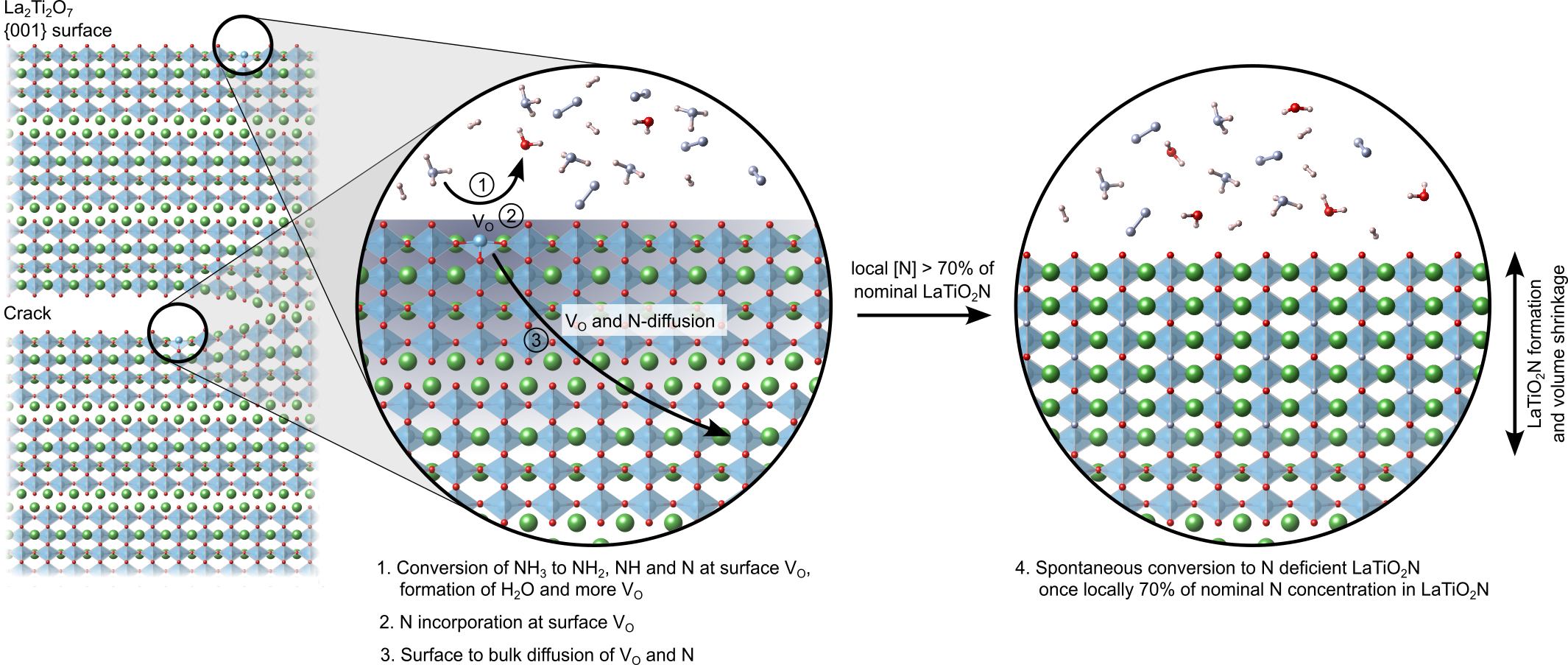}
	\caption{Schematic of the ammonolysis process: Oxygen vacancy defects at surfaces and inside cracks lead to rapid nitrogen incorporation and oxygen removal. Once a critical nitrogen content around 70\% of the stoichiometric \ce{LaTiO2N} concentration is reached, the structure converts into N-deficient \ce{LaTiO2N}. The accompanying volume shrinkage leads to more cracks.}
	\label{fig:schematic}
\end{figure*}

Fig.~\ref{fig:crossover2}a reports the relative energy of the various configurations as a function of their \ce{V_O} location (I, M, B or a mixture thereof), the location of \ce{N_O} and the resulting anion order. This data confirms that what we already learned from isolated defects and defect pairs is also valid at high defect concentrations: \ce{N_O} prefers to be in the middle or bulk layer (low lying red data points) and \ce{V_O} are stabilised in the interlayer when \ce{N_O} are present (lowest energies in the I column). Interestingly, when \ce{V_O} are present in the interlayer, we observe spontaneous zipping of the two LTO slabs during structural optimization, the resulting structure being defective LTON. When zipping is observed, the final optimized structure is energetically slightly more stable than the corresponding LTO cell, as shown in Fig.~\ref{fig:crossover2}c, which reports for each configuration the total energy difference between defective LTON and defective LTO. The most stable defective LTO cell (that spontaneously converts to LTON during relaxation) has a local \textit{trans} anion order. This is in contrast to the most stable \textit{cis} anion order in LTON~\cite{Ninova2017} but is caused by the \textit{trans} order leading to an ideal anion arrangement for zipping, i.e. without nitrogen in the interlayer plane.

We repeated the same calculations for 50\% nitrogen content (8 substitutional N and 4 oxygen vacancies in LTO). In this case, not only do we observed that  \ce{V_O} are more  stable in the bulk and middle layer (see Fig.~\ref{fig:crossover2}b), but, more interestingly, the zipping of the two slabs to form LTON is never observed, LTO always being more stable than LTON (see Fig.~\ref{fig:crossover2}d). The transformation of LTO to LTON is thus hindered at low N content both energetically as well as by the absence of \ce{V_O} in the interlayer that are required to induce zipping.

\section{Conclusions}

Our results indicate that surface \ce{V_O} are crucial for the complete decomposition of \ce{NH3} to N on the LTO surface. While, in principle, the initial decomposition steps could also occur on the stoichiometric surface, the presence of (\ce{H2} induced) \ce{V_O} will lead to clustering of \ce{NH} adsorbates that are essential to reduce the energetic cost of the final decomposition step ultimately leading to N incorporation into LTO. Since bulk \ce{V_O} concentrations are low, due to the rather high formation energy of this defect, \ce{V_O} will have to be created at the surface.

Once incorporated at the surface, N prefers to reside on an O site rather than as an interstitial and to assume a fully ionic \ce{N^{3-}} charge state. This species is more stable within the middle and bulk layer, suggesting that N diffuses away from the surface and into the LTO crystal. While diffusion parallel to the surface is most facile, the barriers for diffusion into the material are surmountable at typical thermal ammonolysis temperatures. This suggests that N will diffuse both laterally and vertically away from the surface and gradually saturate a region below the initial \ce{V_O} defects with substitutional \ce{N_O}. When nitrogen substitutes for an oxygen, the oxygen atom can become an interstitial, which can be removed from the material by diffusion and/or annihilation with oxygen vacancies. The presence of interstitial O is also in line with the experimentally observed anion superstoichiometry prior to conversion to LTON.

Our calculations indicate that once the local nitrogen content reaches about 70\% of the concentration in stoichiometric LTON, the LTON structure becomes thermodynamically more stable than LTO. This structural transformation is spontaneous within our calculations as long as oxygen vacancies are present in the interlayer region, which is the case only for sufficiently high \ce{N_O} concentrations. Based on comparison with experimentally determined UV-vis spectra, the resulting LTON is likely to contain some metastable interstitial N caused by a high local N concentration in the N saturated region below surface \ce{V_O}.

Figure \ref{fig:schematic} schematically shows this process of \ce{NH3} decomposition at \ce{V_O} on the predominant {001} surfaces and crack faces, followed by N and \ce{V_O} diffusion until a local N content $\sim70\%$ of nominal \ce{LTON} is reached and ultimately collapse of the layered \ce{LTO} structure to N deficient \ce{LTON}.

\section*{Author Contributions}
C.R., S.P. and U.A. conceptualised research. C.R., T.B. and X.W. conducted DFT calculations, supervised by U.A., V.W. conducted experiments, supervised by S.P., C.R. and U.A. wrote the original draft, all authors reviewed and edited the manuscript.

\section*{Conflicts of interest}
There are no conflicts to declare

\section*{Acknowledgements}
This research was funded by the Swiss National Science Foundation (SNSF) Professorship Grants PP00P2\_157615 and PP00P2\_187185, SNSF Project 200021\_178791 and the NCCR MARVEL, a National Centre of Competence in Research, funded by the SNSF (grant number 182892). Computational resources were provided by the University of Bern (on the HPC cluster UBELIX, http://www.id.unibe.ch/hpc) and by the Swiss National Super Computing Center (CSCS) under project IDs s955 and s1033.

\bibliography{references}

\end{document}